\documentclass[a4paper, amsfonts, amssymb, amsmath, reprint, showkeys, nofootinbib, twoside]{revtex4-1}
\usepackage[utf8]{inputenc}
\usepackage[colorinlistoftodos, color=teal!40, prependcaption]{todonotes}
\usepackage{amsthm}
\usepackage{mathtools}
\usepackage{physics}
\usepackage{graphicx}
\usepackage[left=15mm,right=15mm,top=35mm,columnsep=15pt]{geometry} 
\usepackage{adjustbox}
\usepackage{placeins}
\usepackage{lipsum}
\usepackage{csquotes}
\usepackage{listings}
\usepackage{amsmath}
\usepackage{float}

\usepackage{wrapfig}

\newcommand{\mb}{\mathbf}
\newcommand{\mc}{\mathcal}

\definecolor{codegray}{rgb}{0.3,0.3,0.3}
\definecolor{bg}{rgb}{0.96, 0.96, 0.96}

\lstdefinestyle{input}{
    backgroundcolor=\color{bg},   
    keywordstyle=\bfseries\color{red!40!black},
    commentstyle=\itshape\color{purple!40!black},
    numberstyle=\tiny\color{codegray},
    stringstyle=\color{purple},
    identifierstyle=\color{red!60!black},
    basicstyle=\ttfamily\footnotesize,
    breakatwhitespace=false,         
    breaklines=true,                 
    captionpos=b,                    
    keepspaces=true,                 
    numbers=left,                    
    numbersep=5pt,                  
    showspaces=false,                
    showstringspaces=false,
    showtabs=false,                  
    tabsize=2
}

\lstdefinestyle{code}{
    backgroundcolor=\color{bg},   
    keywordstyle=\bfseries\color{green!40!black},
    commentstyle=\itshape\color{purple!40!black},
    numberstyle=\tiny\color{codegray},
    stringstyle=\color{purple},
    identifierstyle=\color{blue},
    basicstyle=\ttfamily\footnotesize,
    breakatwhitespace=false,         
    breaklines=true,                 
    captionpos=b,                    
    keepspaces=true,                 
    numbers=left,                    
    numbersep=5pt,                  
    showspaces=false,                
    showstringspaces=false,
    showtabs=false,                  
    tabsize=2
}

\usepackage[pdftex, pdftitle={Article}, pdfauthor={Author}]{hyperref} 
\usepackage{xcolor}

\usepackage{hyperref}
\hypersetup{
    colorlinks=true,
    linkcolor=blue,
    filecolor=blue,      
    urlcolor=blue,
    citecolor=blue,
    pdfpagemode=FullScreen,
    }
    
\begin{document}

\preprint{AIP/123-QED}

\title{MATILDA.FT, a Mesoscale Simulation Package for Inhomogeneous Soft Matter}

\author{\href{https://orcid.org/0000-0001-8127-7658}{\includegraphics[scale=0.06]{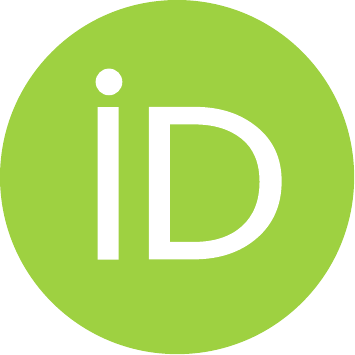}} Zuzanna M. Jedlinska}
\affiliation{Department of Physics and Astronomy, University of Pennsylvania, Philadelphia, USA}

\author{Christian Tabedzki}
\affiliation{Department of Chemical and Biomolecular Engineering, University of Pennsylvania, Philadelphia, USA}

\author{Colin Gillespie}
\affiliation{Department of Chemical and Biomolecular Engineering, University of Pennsylvania, Philadelphia, USA}

\author{Nathaniel Hess}
\affiliation{Department of Chemical and Biomolecular Engineering, University of Pennsylvania, Philadelphia, USA}

\author{Anita Yang}
\affiliation{Department of Chemical and Biomolecular Engineering, University of Pennsylvania, Philadelphia, USA}

\author{\href{https://orcid.org/0000-0002-5434-4787}{\includegraphics[scale=0.06]{img/orcid.pdf}}Robert A. Riggleman}
    \email[Correspondence email address: ]{rrig@seas.upenn.edu}
    \affiliation{Department of Chemical and Biomolecular Engineering, University of Pennsylvania, Philadelphia, USA}

\date{\today} 

\begin{abstract}

In this paper we announce the public release of a massively-parallel, GPU-accelerated software, which is the first to combine both coarse-grained molecular dynamics and field-theoretical simulations in one simulation package. MATILDA.FT (Mesoscale, Accelerated, Theoretically-Informed, Langevin, Dissipative particle dynamics, and Field Theory) was designed from the ground-up to run on CUDA-enabled GPUs, with the Thrust library acceleration, enabling it to harness the possibility of massive parallelism to efficiently simulate systems on a mesoscopic scale. MATILDA.FT is a versatile software, enabling the users to use either Langevin dynamics or Field Theory to model their systems - all within the same software. It has been used to model a variety of systems, from polymer solutions, and nanoparticle-polymer interfaces, to coarse-grained peptide models, and liquid crystals. MATILDA.FT is written in CUDA/C++ and is object oriented, making its source-code easy to understand and extend. The software comes with dedicated post-processing and analysis tools, as well as the detailed documentation and relevant examples. Below, we present an overview of currently available features. We explain in detail the logic of parallel algorithms and methods. We provide necessary theoretical background, and present examples of recent research projects which utilized MATILDA.FT as the simulation engine. We also demonstrate how the code can be easily extended, and present the plan for the future development. The source code, along with the documentation, additional tools and examples can be found on GitHub \href{https://github.com/rar-ensemble/MATILDA.FT}{MATILDA.FT} repository.

\end{abstract}


\maketitle

\section{Introduction}

Poymers are a ubiquitous type of material, important both in biological and industrial settings. Polymers are an umbrella term, gathering macromolecules composed of smaller, repeating monomers. In the industrial settings, polymers are used extensively in the tire industry, in flexible composite material production, and utilized as durable adhesives. In addition, they have been exploited in more precise applications, such as drug delivery \cite{indulkar_exploiting_2016} and the design of artificial catalysis centers \cite{sureka_catalytic_2019}. This is possible due to the propensity of polymers to self-assemble into higher-order structures, and their ability to undergo phase separation in solution. Controlled phase separation has been exploited to create nano-capsules with well-defined pore sizes, by first inducing phase separation in the capsule shell and then flushing-out one of the components \cite{shi_polymer_2017}. Similar approaches using non-solvents to induce phase separation in a polymer solution are common methods to produce polymer membranes\cite{haase2017multifunctional, tree2017multi}. Design of these materials requires precise knowledge of the thermodynamics and microstructure of polymer materials under a variety of conditions.

Within a biological context, the most important natural polymers are nucleic acids and proteins. The former are responsible for storing and propagating the genetic information, whereas the latter act as enzymes to facilitate multiple biochemical reactions within the cell, participate in active transport of intracellular components, or serve as structural elements of the cytoskeleton. A special class of proteins, called Intrinsically Disordered Proteins (IDPs) have been shown to be a main constituent of the membraneless organelles \cite{wei_phase_2017}. The functions of these intracellular compartments vary from nucleic-acid biosynthesis and organization, to participation in stress, and immune responses \cite{banani_biomolecular_2017}. Just like regular organelles, they have a distinct chemical composition, and thus spatially separate specific biological processes. Unlike regular organelles, however, they are not enclosed by a lipid membrane, and thus membraneless organelles can be easily formed and dissolved as the need arises. The deregulation of these process has been implicated in cancer development and the onset of neurodegenerative disease \cite{kanaan_liquid-liquid_2020}. Efficient prediction of the equilibrium thermodynamics of these systems remains an ongoing challenge, although significant recent progress has been made\cite{choi_physical_2020}.

Given their wide-ranging applications, inhomogeneous (macro- or micro-phase separated) polymers remain a topic of intense investigation. Experimental methods are available to study their properties and aid in the novel material design, and the polymer science community has a rich history of close collaboration between experiment, theory, and simulation. Computer-aided approaches have had an ever increasing importance as a research tool as computational power has become more ubiquitous. Due to this continuously increasing computational speed and processing power, progressively bigger and more complex systems can be simulated, allowing mesoscopic material propertiers to be predicted before synthesizing physical samples.

Various open-source Molecular Dynamics (MD) codes have been released, which are capable of simulating polymeric species on an atomistic or coarse-grained level. Some notable examples include LAMMPS \cite{thompson_lammps_2022}, NAMD \cite{phillips_namd_2022}, and GROMACS \cite{abraham_gromacs_2015}. LAMMPS can perform all-atom simulations on polymer chains, using available force fields. It is also equipped with biologically-oriented force fields, which enable coarse grained simulations of biomolecules. In addition, the user can define their own coarse-grained polymer model, and expand it to include the required potentials. On the other hand, both GROMACS and NAMD have been specifically designed to model biological molecules, such as proteins and nucleic acids on a fully atomistic level. Polymer simulations are challenging in general due to the wide range of length- and time-scales required for accurate simulation. 

In many soft matter fields, particularly those involving the design materials using polymers, polymer field theory and related techniques have played a crucial role in the design of new materials and in the interpretation of experimental results\cite{ETIP, edwards1965statistical, leibler1980theory, zheng1995morphology, milner1988theory}. Polymer field theories are developed by beginning with a description of the system in terms of coarse-grained potentials, such as chains obeying Gaussian statistics, Flory contact repulsions governed by a $\chi$ parameter, partial charges on the various species, etc. One first writes down the partition function for this particle model, then using one of a variety of transformation techniques\cite{ETIP, edwards1970cross, edwards1965statistical}, decouples the particle interactions and transforms the model to one where one molecule of each type interacts with chemical potential fields generated by the various interaction potentials. The field-theoretic approach is attractive because it enables a variety of analytic analysis, such as the mean-field approximation which gives rise to self-consistent field theory (SCFT) or a variety of loop expansions. The particle-to-field transformation is formally exact, and there are examples in the literature showing quantitative agreement between the particle and field version of the model\cite{koski2013field}. In more recent years, several methods have developed to sample the original particle model efficiently\cite{detcheverry2008monte, pike2009theoretically, daoulas2006single, ganesan2003dynamical, chao2017solvent} that generally fall under the umbrella of theoretically-informed coarse-grained models; in these methods, the underlying particle coordinates are retained, and the particles are mapped to density fields to efficiently calculate the non-bonded forces and energies. However, with a few notable exceptions\cite{schneider2019multi, arora2016broadly, cheong2020open}, there is a dearth of simulation packages available to perform general simulations on either particle or field-based models, particularly those that are designed specifically to run on Graphics Processing Units (GPUs). A single code base that can simulate both particle- and fluctuating field-theoretic simulations of identical (or very similar) molecular models could allow readily switching between dynamic and equilibrium simulations as well as assessment of the importance of fluctuations in any calculating properties.

In this work, we present a first version of our code for Mesoscale, Accelerated, Theoretically-Informed, Langevin, Dissipative particle dynamics, and Field Theory, MATILDA.FT. MATILDA.FT is written from the ground-up intended to be run on GPUs, and the bulk of the code is written using the CUDA programming language. MATILDA.FT is capable of modeling both the systems consisting of a few molecules, as well as those containing millions of particles. Its strength lies specifically in being able to efficiently simulate polymeric and other soft materials (e.g., liquid crystalline systems) on a mesoscopic scale. On this scale, the coarse-grained interactions are typically ``soft'' (finite at overlap) and the particle density high; in this limit, it becomes more efficient to evaluate the non-bonded interactions using density fields. These large-scale molecular assemblies of interest can correspond to biomolecular coacervates in explicit solvent, artificially synthesized ionomers, block copolymer melts, side-chain liquid crystalline polymers, or polymer-infiltrated nanoparticle packings. 

The outline of this paper is as follows. In Section \ref{sec:history} we begin with a brief history of the code development. Next, in Section \ref{sec:feature overview} we outline the main features of the code and available functionalities. In Section \ref{sec:models}, we describe the structure of the models being used in the molecular dynamics and field-theoretical simulations. Subsequently, in Section \ref{sec:algorithms}, we show how the code is optimized for parallel execution on CUDA-enabled GPUs. In the Methods Section \ref{sec:methods} we present a more in-depth description of selected features available in MATILDA.FT. In Section \ref{sec:code structure} we outline how the code is organized, with focus on its class structure, and extensibility options. Next, in Section \ref{sec:examples} we show results for selected example systems. We end with the planned developments in Section \ref{sec:future dev}, and a conclude with Section \ref{sec:summary}.

\subsection{Brief History}
\label{sec:history}

The GPU-TILD code evolved out of a series of programs originally referred to as dynamic mean-field theories (DMFT) developed by Chao, Koski, and Riggleman\cite{chao_solvent_2017} and augmented by several students in subsequent years. Over time, it became clear that the approach was strongly connected to existing methods that went by the moniker theoretically-informed Monte Carlo (TIMC). The technical differences between the two is that our approach tends to use higher-order particle-to-mesh schemes to ensure UV-convergence and that time evolution is through Langevin dynamics in lieu of Monte Carlo dynamics. As a result we have subsequently referred to the method as theoretically-informed Langevin dynamics (TILD).Our group had several versions of an internal code that was developed using openMP and later openMPI. However, all of these codes suffered from a fairly rigid structure that was tied to simulations of specific systems associated with specific research projects and an unintuitive input file format.

During the first summer of the Coronavirus pandemic, RAR began developing a basic TILD simulation package. The goal was to develop code intended to be used on GPUs from the ground up, creating a more modular, object-oriented structure. Once the utility of the code became clear, its use spread throughout our research group, with substantial additions to the code base being made by all of the co-authors of this work, most notably CT and ZMJ.
The addition of the field-theoretic simulations portion of the code began in late 2022 by RAR.

\section{Feature overview}
\label{sec:feature overview}

In this section, we provide a brief overview of the features available in MATILDA.FT. They are later described in more detail in the following sections, \ref{sec:algorithms} and \ref{sec:methods}.

MATILDA.FT can performed two types of simulations: TILD or field-theoretic (FT) simulations. The TILD method is a hybrid particle/field approach where explicit coordinates of the molecules are retained and used to calculate bonded interactions while non-bonded forces (e.g. excluded volume and electrostatics) are calculated by mapping the particles to a density field. The highly coarse-grained nature of the interactions leads to a significant speed-up compared to particle-only implementations of the same models, due to the relatively high particle density in such models. The method is closely related to the theoretically-informed coarse-grained modeling techniques developed in the de Pablo group\cite{detcheverry2008monte, pike2009theoretically} and other related techniques\cite{daoulas2006single, ganesan2003dynamical}. In a FT simulation, on the other hand, particle coordinates are integrated out completely and one must compute the statistics of the molecular conformations in external fields generated by the other molecules. FT simulations are especially powerful when the equilibrium properties of high molecular weight polymers are of interest, specifically a large dimensionless polymer concentration $C = n / (V/R_g^3)$, where $n$ is the number of chains in a system, $V$ is the volume, and $R_g$ is the radius of gyration of a chain. The FT simulations and TILD methods are complimentary in this sense, as the particle-based TILD approach is much more efficient at lower polymer concentrations, and TILD can capture some aspects of polymer dynamics that FT cannot.

\subsection*{\textbf{TILD Branch}}

Next, we move to the overview of the TILD module. Here, the simulations are performed in the NVT ensemble, in a fully-periodic orthogonal box, either in two or three dimensions.
Although MATILDA.FT can perform simulations of free particles, it has been designed specifically to efficiently model systems of polymer melts and solutions. Polymers are modeled as discrete Gaussian chains with monomers that are connected through harmonic springs, and the density of each monomer is spread around its center through a convolution with a unit Gaussian. The strength of the repulsive interactions between the species is mediated through the Flory-Huggins $\chi$ parameter. Monomers can be either neutral or charged. If they carry a net charge, then in addition to the repulsive potential, they also interact through Coulombic electrostatic forces. Regardless of their net charge, the monomers can be made polarizable through the use of (classical) Drude Oscillators as detailed in Section \ref{subsec:Drude oscillator}.

User-defined groups of particles are the basic structure to which operations in MATILDA.FT are assigned. These operations can be either static or dynamic. Static groups maintain the same set of members over the entire course of simulation whereas dynamic groups periodically update their member lists, based on a specific membership criterion. The basic set of operations acting on particle groups are applying the repulsive and electrostatic force, thermostating, and integration, in order to propagate coordinates in time. Other operations that can be applied to particles include external/biasing forces, spatial confinement, creation and breaking of dynamic bonds, or performing on-the-fly property calculations. Some of these operations require a distance-based neighbour list, MATILDA.FT provides different styles of neighbour lists depending on the application.

The user interacts with the code through input scripts, written in a plain-text format. Before the simulation is started, the entire script is read and appropriate variables and data structures are initialized. The maximum number of time steps and the time step size are parameters specified in the input script. Then the system can undergo an optional equilibration period before subsequently entering the production run stage. 
Different input formats are required by the TILD and FT simulations. For the TILD branch, two files need to be provided. The first one is the main input file, providing information about simulation dimensionality, box size, density grid spacing, and interaction potentials between particle types. It also defines the particle groups and assigns integrators and forces to act on them during each time step.
The second file contains information about particle coordinates, types, molecules they belong to, and optionally, their charge. Currently, this file can be provided in the format consistent with the LAMMPS data file, in either \textit{atomic} or \textit{charge} atom style. To allow the use of data generated by other codes, the initial configuration can also be read from a GSD-format file, developed by the Glotzer Lab \cite{glotzer_lab_gsd_2023}. On the other hand, FT simulations only require a single input file, as all the fields and molecular information are initialized from within the input script.

A sample input script, providing basic TILD functionality is shown in Listing \ref{infile}. Here, the simulation is performed in 2 dimensions, for 10,000 time steps. The script specifies various log and output frequencies. The variable \textit{pmeorder} controls the order of interpolation for constructing the density fields. The dimensionless time step is set to 0.005. The configuration data is read from the \textit{input.data} file. An external force \textit{midpush} (along the y-axis, with the dimensionless magnitude of 0.5) is applied to all particles. The integrator is chosen to be GJF. The lines beginning with \textit{pair\_style} specify the repulsive interactions between the selected particles types. A more detailed description of all available options can be found in the documentation and in Section \ref{sec:code structure}.

\begin{lstlisting}[language=C++,caption={Example input script} ,captionpos=b, label=infile, style = input]
Dim 2

max\_steps 10001
log_freq 1000
binary_freq 1000
pmeorder 1
traj_freq 1000

delt 0.005
read_data input.data

Nx 65
Ny 65

extraforce all midpush 1 0.5

integrator all GJF

pair_style gaussian 1 1 1.5625  1.0
pair_style gaussian 2 2 1.5625  1.0
pair_style gaussian 1 2 3.00  1.0

\end{lstlisting}

\subsection*{\textbf{FT Branch}}

As the development of the FTS features of the code only began relatively recently, the feature set is currently somewhat limited, but will expand significantly in the coming months. Currently, FTSs are limited to mean-field calculations as in self-consistent field theory (SCFT) with linear, discrete Gaussian chain models. The molecules can be of arbitrary blockiness with an arbitrary number of components, and the potentials implemented include the Flory contact repulsion and the Helfand weak compressibility\cite{helfand_theory_1975}. More details about the interactions between the species are provided in Section \ref{sec:models}. As detailed below, the key elements of the FTS implementation are three classes: {\bf Potentials}, which govern the non-bonded interactions and act on {\bf Species}. The {\bf Species} class stores the total density of each chemical component and is populated by individual {\bf Molecule} classes. For example, an A-homopolymer/B-homopolymer/AB-diblock copolymer blend would have two species (A and B) and three molecules.

\subsection*{Units}

All simulations in MATILDA.FT are performed in reduced units. The base units for TILD simulations are \textit{energy} $[k_bT]$, \textit{mass} $m$ of a monomer, and \textit{length} is normalized by the statistical segment size $b$. For the FT simulations, the unit length scale is the radius of gyration computed for an ideal chain of a specified reference chain length, $N_r$, $R_g = \sqrt{(N_r-1)/6}$. In addition, the field variables are also scaled by this reference chain length so that the natural interaction parameters become $\chi N_r$, $\kappa N_r$, etc. By substituting an appropriate value, other derived units can be obtained, such as time and force. 

\subsection*{Output and Additional Features}

The user might want to retrieve the information generated during the simulation. By default, a frame of the trajectory file is written with a specified frequency. This trajectory file is consistent with the LAMMPS traj file style. Since this operation slows down the program execution, an alternative binary file can also be saved. Two binary outputs are written out - one storing particle positions and another storing the density fields. These can be later converted to a human-readable format with the post-processing tools that are distributed with the code.
The simulation can be initialized from a configuration file, following LAMMPS \textit{atomic} or \textit{charge} input styles. It can also be resumed from the previously generated trajectory file by using a resume command in the input script.
Each job run can also be split into an equilibration and production sections by explicitly setting the number of steps for each section in the input script. This writes the data to two separate files, each of which can have their write rates modified independently of other sections. This allows for data to be collected during and post equilibration within a single run, without a need to manually resume the simulation. The trajectory data can also be saved in the GSD format. More details about the specific output options can be found in thedocumentation.

\section{Structure of theoretically-informed coarse-grained and field-theoretic models}
\label{sec:models}

In this section we provide the necessary theoretical background to understand the models used in MATILDA.FT, and the logic of the simulation workflow. The starting ingredients for all of the modeling to be handled by MATILDA.FT are highly coarse-grained models for soft-matter systems. For simplicity, we will describe the basic structure in terms of a simple A-B Gaussian chain diblock copolymer melt, though the generalization to other systems will become apparent below. For a polymer melt with $n$ polymer chains each containing $N_A + N_B = N$ monomers, the microscopic polymer densities are
\begin{equation}
    \hat \rho_K(\mb{r}) = \sum_j^{n} \sum_s^{N_K} \delta (\mb r - \mb r_{j,s}),
    \label{eq:rhohat}
\end{equation}
where $K$ is either species A or B, and $\mb r_{j,s}$ is the position of the $s^{\mathrm{th}}$ bead on the $j^{\mathrm{th}}$ chain. The monomers on each chain are typically connected via harmonic bonds,
\begin{equation}
    \beta U_0 = \sum_j^n \sum_s^{N-1} \frac{3}{2b^2} |\mb{r}_{j,s} - \mb r_{j,s+1}|^2,
    \label{eq:bonds}
\end{equation}
where $b$ is the statistical segment size, and we have assumed equal $b$ for species A and B. The Flory repulsion is written in one of two equivalent forms\cite{koski2013field, villet_efficient_2014, weyman2021field} depending on whether the model is implemented as a field- or particle-based model. In the particle-based approaches, we make the potential non-local as
\begin{equation}
    \beta U_1 = \frac{\chi}{\rho_0} \int d\mb{r} \int d\mb{r}' \, 
    \hat \rho_A(\mb r) 
    \, 
    u_G(|\mb r-\mb r'|)
    \,
    \hat \rho_B(\mb r'),
    \label{eq:fh}
\end{equation}
where $u_G$ is a unit Gaussian potential, $u_G(r) = (2\pi\sigma^2)^{-\mathbb{D}/2} e^{-r^2/2\sigma^2}$, $\mathbb{D}$ is the dimensionality of the system, and $\sigma$ controls the range of the interactions. The standard Flory-Huggins model is recovered in the limit $\sigma\rightarrow 0$. 
The final potential penalizes deviations of the local density from the average\cite{helfand1975theory} $\rho_0 = nN/V$,
\begin{equation}
    \beta U_2 =  \frac{\kappa}{2\rho_0} \int d\mb r \int d\mb r' \; [\hat \rho_+(\mb r)-\rho_0] \, u_G(\mb r - \mb r') \,  [\hat \rho_+(\mb r')-\rho_0]
    \label{eq:HelfandP}
\end{equation}

With these ingredients in hand, we can write the partition function as
\begin{equation}
    \mc Z = 
    z_0 \int d\mb r^{nN} \; e^{-\beta U},
    \label{eq:Zp}
\end{equation}
where $z_0$ is a prefactor that contains all of the self-energy terms, factors accounting for molecular indistinguishability, and the thermal de Broglie wavelengths. In equilibrium particle-based simulations, one is primarily interested in calculating averages of quantities that can be expressed as functions of the particle coordinates, $M(\mb r^{nN})$, as
\begin{equation}
    \langle M \rangle = \frac 1 {\mc Z} \int d\mb r^{nN} \, M(\mb r^{nN}) \, e^{-\beta U},
\end{equation}
and expressions for the usual thermodynamic quantities of interest, such as the average density, energy, and pressure, can be readily obtained from expressions commonly used in molecular dynamics simulations\cite{allen2017computer}.

For field-theoretic approaches, we use a local potential but render the densities non-local by distributing the point particles over a Gaussian distribution $h(r)=(2\pi a^2)^{-\mathbb{D}/2} e^{-r^2/2a^2}$, and the Gaussian-distributed particle density is given by 
\begin{equation}
    \breve \rho_K(\mb r) = \int d\mb r' \; h(\mb r - \mb r') \, \hat \rho_K(\mb r')
    = [h \ast \hat \rho_K](\mb r),
    \label{eq:smearRho}
\end{equation}
where the final equality introduces our short-hand notation for a convolution integral. For the choice $\sigma^2 = 2a^2$, we can exactly re-write\cite{koski2013field, villet_efficient_2014, weyman2021field} the non-bonded potentials in Eqs.~\ref{eq:fh} and \ref{eq:HelfandP} as
\begin{equation}
    \beta U_1 = \frac{\chi}{\rho_0} \int d\mb r \; \breve \rho_A(\mb r)\, \breve \rho_B(\mb r),
    \label{eq:fhF}
\end{equation}
and
\begin{equation}
    \beta U_2 = \frac{\kappa}{2\rho_0} \int d\mb r \; [\breve \rho_+(\mb r)-\rho_0]^2.
    \label{eq:HelfandF}
\end{equation}
Using known Gaussian functional integrals \cite{edwards1965statistical, ETIP}, one can then exactly transform the particle-partition function in Eq.~\ref{eq:Zp} to a field-theoretic one of the form
\begin{equation}
    Z = z_1 \int \mc D\{w\} \; e^{-\mc H[\{w\}]},
    \label{eq:Zft}
\end{equation}
where $z_1$ contains the constants from $z_0$ as well as the normalizing factors from the Gaussian functional integrals, $\{w\} = \{w_+, w_{AB}^{(+)}, w_{AB}^{(-)}\}$ is the set of chemical potential fields, and $\mc H$ is the effective Hamiltonian governing the weights of the microstates. For the diblock copolymer model considered here, $\mc H$ takes the form
\begin{equation}
    \begin{array}{rcl}
    \mc H &=& \frac{C}{\chi N_r } \int d\mb r \left( [w_{AB}^{(+)}(\mb r)]^2 + [w_{AB}^{(-)}(\mb r)]^2 \right)
    \\
    && + \frac{C}{2\kappa N_r} \int d\mb r [w_+(\mb r)]^2 - i C \int d\mb r \; w_+(\mb r)
    \\
    && - n_D \log Q_D[\mu_A, \mu_B],
    \end{array}
    \label{eq:H}
\end{equation}
where the first line contains the potential fields that arise due to the Flory interaction\cite{koski2013field}, the second line contains the terms that arise from the Helfand potential, and the final line contains the excess chemical potential of the polymers in a given field. The potential fields $\mu_A$ and $\mu_B$ experienced by monomers A and B computed using
\begin{equation}
    \begin{array}{rcl}
        \mu_{A,c}(\mb r) & = & \left\{i(w_+ + w_{AB}^{(+)}) - w_{AB}^{(-)}\right\} (\mb r) / N_r
        \\ && \\
        \mu_{B,c}(\mb r) & = & \left\{i(w_+ + w_{AB}^{(+)}) + w_{AB}^{(-)}\right\}(\mb r) / N_r,
    \end{array}
    \label{eq:MuKc}
\end{equation}
with the smeared potential fields appearing in Eq.~\ref{eq:H} calculated as $\mu_K(\mb r) = [h \ast \mu_{K,c}](\mb r)$.

While the particle implementation can report qualitatively realistic dynamic quantities, the FT implementation is strictly interested in equilibrium quantities. Equilibrium averages are typically expressed as functionals of the potential fields and calculated as 
\begin{equation}
    \langle M \rangle = \frac 1 {\mc Z} \int \mc D \{w\} \, M[\{w\}] \, e^{-\mc H}.
\end{equation}
Since $\mc H$ is typically complex-valued, sampling the integral over the field configurations is non-trivial; this is typically accomplished through the mean-field approximation, leading to SCFT, through complex Langevin (CL) sampling\cite{ETIP, parisi1983complex, klauder1984coherent}, or Monte Carlo sampling\cite{stasiak2013monte}. 

To update the chemical potential fields in either a CL or an SCFT calculation, the effective ``forces'' on the fields must be obtained as functional derivatives of $\mc H$. These can be obtained through explicit differentiation,
\begin{equation}
    \mb F_w(\mb r) = - \frac{ \delta \mc H } { \delta w(\mb r)},
\end{equation}
where $w(\mb r)$ is one of the three fields $w_+(\mb r), w_{AB}^{(+)},$ or $ w_{AB}^{(-)}$. In a CL simulation, the fields are sampled using an overdamped Langevin equation,
\begin{equation}
    \frac{ \partial w}{\partial t} = \lambda_w \mb F_w(\mb r) + \eta(\mb r, t),
    \label{eq:CL}
\end{equation}
where $\eta(t)$ is a stochastic noise term chosen to satisfy the fluctuation-dissipation theorem\cite{ETIP, lennon2008numerical}. An algorithm that drives the system to an SCFT solution is easily obtained from Eq~\ref{eq:CL} by simply setting the noise term to zero.

\section{Parallel Algorithms, Data Structures and Object-oriented Programming with CUDA and C++}
\label{sec:algorithms}

The unique feature which sets MATILDA.FT apart from other popular MD codes, such as LAMMPS or NAMD is that it has been designed from the beginning to execute on CUDA-enabled GPUs. The code is intended for highly coarse-grained models where the non-bonded forces can be evaluated using density fields and not summing over neighboring pairs of particles. It uses a dedicated CUDA/C++ programming language in order to fully harness parallel capabilities. Its model of parallelization differs from the conventional CPU domain decomposition. Whereas on the CPU the groups of particles are assigned to different processor based on their spatial arrangement, the GPU parallelization occurs on the particle or individual grid location level, where each thread is responsible for processing instruction for the selected particle/grid point. It is simply handled by assigning a separate thread to individual particles, by filtering the thread IDs. This is handled by the first statement in each kernel call, provided in Listing \ref{code:select id}.

\begin{lstlisting}[language=C++, style = code, label = code:select id, caption = Kernel call to assign a thread to a particle id.]
int list_ind = blockIdx.x * blockDim.x + threadIdx.x;
if (list_ind >= ns)
    return;
\end{lstlisting}

MATILDA.FT also makes extensive use of Thrust library, which is an extension of C++ Standard Template Library (STL) to work with GPUs \cite{bell_thrust_2012}. The Thrust Library provides dedicated storage containers (equivalent to STL vectors in C++), which enable easier host-device communication and avoiding the requirement for explicit \textit{cudaMemcpy} calls. The Thrust Library also makes available dedicated parallel algorithms to operate on these containers and achieve better performance. In addition, by avoiding complicated host-device memory transfer syntax, the use of thrust makes it easy for those who do not have much GPU-programming experience to understand and expand the MATILDA.FT source code.

\subsection{Neighbour lists}
\label{subsec:n-lists}

Neighbour lists keep track of the other particles present within a certain distance of the center of the particle of interest. This information is required, for example, when model reactive particles need to search for reaction partners. In this way, each time the force is being applied, a comparison of all possible inter-particle distances, an $\mathcal{O}(N_{tot}^2)$ operation, where $N_{tot}$ is the total number of particles, is avoided. Instead, an $\mathcal{O}(N)$ scaling is achieved during each function call, since only pre-computed neighbours are checked for possible interactions. The process of creating the neighbour list is based on cells, to which the simulation box is divided, and thus it also scales as $\sim N$.

Each neighbour list requires two parameters: $r_{cutoff}$ and $r_{skin}$. They correspond to the maximum range of the force using the neighbour list, and the skin radius $(r_{skin} > r_{cutoff})$ respectively.
The skin radius is the actual radius of the sphere (or a disc in 2D) within which the search for the possible neighbours occurs. By specifying the skin radius, some particles which are initially beyond the range of the force are included in the neighbour list, and thus the list can be reused over multiple time steps. They neighbour list can be rebuilt with a specified frequency, or, alternatively, an automatic trigger can be used. Currently, list rebuilding is triggered  when a particle moves a distance larger than half the skin radius.

Each cell is chosen to have a side length equal to the half of the specified skin radius.
The process of neighbour list building is divided into two stages. In the first phase, particle "binning" takes place, where each particle in the group gets assigned to one cell in the box. Finding the corresponding cell is trivially parallelized over all particles, scaling as $\mathcal{O}(N)$, whereas their assignment the grid point requires an atomic operation. Each cell stores the ID of its member particles, and keeps track of their total count.
Subsequently, the program loops over each particle in the group and checks the distance to other particles present in the cells overlapping with $r_{skin}$. Each cell is chosen to have a side length equal to the half of the specified skin radius, resulting in 125 cells in 3D, and 25 in 2D. Only the particles within the specified radius are included as particle's neighbours. The full neighbor list is stored as a 1D thrust device vector, so no host-GPU data transfer is required. Another 1D device vector stores cell content and their respective occupation count. This method uses a predefined maximum cell capacity. If the specified capacity is too small, it gets automatically adjusted during the simulation run. The full neighbour list stores all neighbours of each particle, effectively double counting each particle pair. A "half-list" stores, for each particle, only the neighbours with index lower than its own index. Other, specialized lists are also available, and interface with specific particle-particle operations. For example, to aid in dynamic bonding, where each donor particle only stores the information about the acceptors present within the skin radius.
In the future release, other, more efficient schemes of neighbour lists will be implemented. For example, each particle can be responsible for the region covering only half of its neighbourhood, such that when combined, the particles cover the search space in a minimally overlapping manner. 

\subsection{Bonded interactions}

Bonded interactions are present whenever polymer chains are modeled. Currently, they represent the harmonic springs connecting adjacent monomers of the same molecule. The calculation of the resulting forces has been parallelized to be performed on the GPU, and Newton's third law is {\it not} used to avoid the use of atomic operations. The contribution of the bonded interactions is calculated individually for each particle by assigning it to a separate tread. MATILDA.FT can also report the resulting bonded energy and the contribution the bonded interactions make to the pressure virial coefficient. These calculations are performed on the CPU. 

Two common angle potentials are also implemented in MATILDA.FT. To enable simulations of discrete worm-like chains, we have the cosine form
\begin{equation}
    u_{wlc}(\theta_{ijk}) = \lambda \left[1 + \cos(\theta_{ijk})\right],
    \label{eq:wlc}
\end{equation}
where $\lambda$ controls the stiffness of the potential and $\theta_{ijk}$ is the {\it inside} angle between particles $i, j$ and $k$. The second potential implements harmonic angles as
\begin{equation}
    u_{h}(\theta_{ijk}) = k_\theta \left( \theta_{ijk} - \theta_0 \right)^2,
\end{equation}
with spring constant $k_\theta$ and equilibrium angle $\theta_0$. The angle styles are specified in the input script along with the type (``wlc'' or ``harmonic''), followed by the force constant (for both styles) and the equilibrium angle if harmonic angles are used.

\subsection{Long-range interactions}

\begin{figure}
    \centering
    \includegraphics[width=0.5\textwidth]{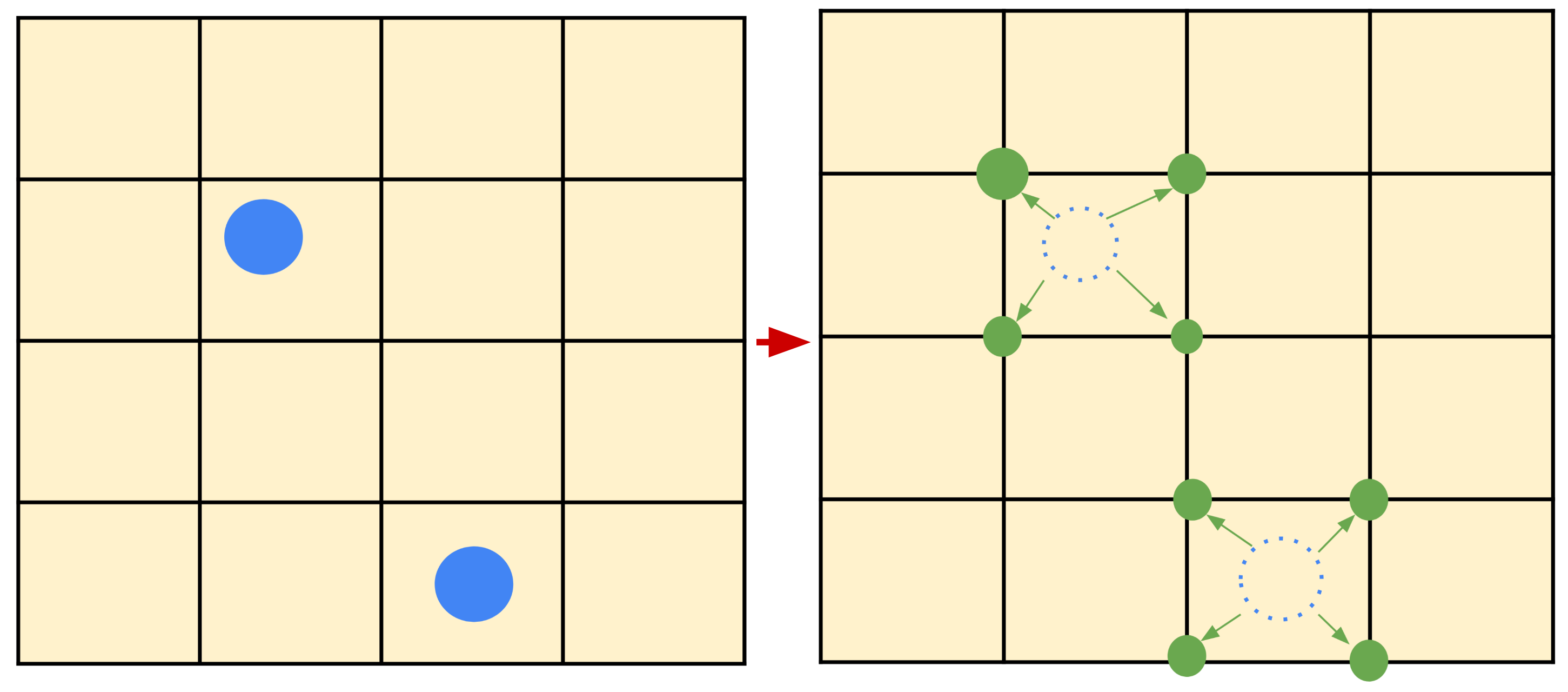}
    \caption{Schematic illustration of the Particle-to-Mesh (PM) scheme. Specifically, shown here is a first-order spline interpolation, where the particle density is mapped to the two nearest grid points in each dimension. The same spline weights are used to map the forces back to the particles.}
    \label{fig:pmm}
\end{figure}

Long-range interactions include repulsive interactions mediated by the Flory-Huggins $\chi$ parameter, and the electrostatic forces acting between the charged monomers. The distinctive feature of MATILDA.FT is the way in which it handles these interactions. While bonded interactions use explicit coordinates to calculate inter-particle distances, long-range interactions use the mass/charge density field to compute the resulting forces. In this process a Particle-to-Mesh (PM) method is used. In this scheme, the box is divided into a discrete grid, with the number of grid-points in each direction being a user-defined quantity. At the beginning of the simulation, a Fourier-space representation of the inter-particle potential is calculated using Fast Fourier Transform (FFT), and is stored for the rest of the simulation. Then, at each time-step, every particle assigns its density contribution to nearby grid points, using a spline interpolation scheme with the weights given in the appendix of Ref.~\cite{deserno_how_1998}. The order of the interpolating spline can be chosen from 1 to 4, with higher order interpolation requiring more computation. Regardless of the form of the pair potential $u(r)$, forces are given in real space by
\begin{equation}
\label{eq:PMM}
    f(\mb r) = - \int d\mb r' \nabla u(\mb r - \mb r') \rho(\mb r').
\end{equation}
The convolution in Eq.~\ref{eq:PMM} is evaluated in Fourier space using FFTs, where it becomes a simple multiplication. Then an inverse FFT is used to transform the forces back into real space where the forces are interpolated back onto the particle centers. The process is illustrated schematically in Figure \ref{fig:pmm}. The current pair potentials implemented in MATILDA.FT include Gaussian forms as well as the nanoparticle-nanoparticle and nanoparticle-monomer forms demonstrated in previous work by some of us\cite{koski2013field}. 

\subsection{Liquid Crystalline Systems}

We model liquid crystalline interactions through a modified Maier-Saupe (MS) potential that is a discrete version of the McMillan model\cite{mcmillan1971simple}. In our implementation, the MS interactions involve two pairs of particles: one of each pair is the ``center'' of the interaction $i$, and the other becomes a partner particle $j$ that is used to define the local molecular orientation on particle $i$, $\mb u_i = \frac{\mb r_i - \mb r_j}{|\mb r_i - \mb r_j|}$ (see schematic in Fig. \ref{fig:LC} below). The local orientation vector is used to define an orientation tensor for each particle $\mb S_i = \mb u_i \mb u_i - \mb I/\mathbb{D}$, which is mapped onto an orientation field similar to the density fields,
\begin{equation}
    \mb S(\mb r) = \sum_{i} \mb S_i \, \delta (\mb r - \mb r_i).
    \label{eq:LC_S_map}
\end{equation}
In Equation \ref{eq:LC_S_map}, the sum is over the LC centers. The orientation field $\mb S(\mb r)$ is then used to compute the Maier-Saupe potential energy\cite{pryamitsyn2004self},
\begin{equation}
    \beta U_{MS} = 
    - \frac{\mu}{\rho_0} \int d\mb r \int d\mb r' \; \mb S(\mb r) :\mb S(\mb r')\, u_G(|\mb r-\mb r'|),
    \label{eq:MSU}
\end{equation}
where $\mu$ is the Maier-Saupe potential parameter and $u_G(r)$ is the Gaussian potential that renders the interactions non-local. The forces are derived by explicit differentiation and are presented in the documentation of the code, and as we show below in Section \ref{sec:examples}, this model captures both nematic and smectic A phases.

Particles that carry an orientation vector $\mb u_i$ are specified in an additional input file that is similar in nature to the lists of bonded partners. When specifying that the MS potential is to be used, the name of the additional input file is also provided; contained in this file is a list of pairs of particles $i$ and $j$ used to define the orientation vector associated with particle $i$. This implementation allows for easy creation of either main-chain liquid crystalline polymers or side-chain liquid crystalline polymers, a detailed study of which will be the subject of a forthcoming publication. Furthermore, by making one of the end sites within an LC mesogen a different site type, one can indirectly control anchoring conditions at phase boundaries by making making this other type more or less repulsive with a particular species in the nearby phase.

\section{Methods}
\label{sec:methods}

In this section, we describe in more detail how selected functionalities are implemented in MATILDA.FT. When possible, we emphasize how the algorithms presented below were optimized to take advantage of the GPU architecture to accelerate performance. The summary of all available functionalities and parameters can be found in the documentation. 

\subsection{Drude oscillators}
\label{subsec:Drude oscillator}

Polarization can play an important role in phase behaviour of polymer solutions, especially in biological context. Molecular polarizability has influence on polymer solubility, and it can also modify how polymer chains interact with salt ions present in the environment. In MATILDA.FT, polarizability effects are introduced through the use of classical Drude oscillators. In this approach, a ``Drude particle'' is attached to the parent particle via a harmonic spring with stiffness $k_D$ and zero equilibrium length. This Drude particle is assigned a partial charge $\delta q_D$ and the partner particle $-\delta q_D$, such that the net charge of the two-particle pair remains unchanged. The magnitude of the spring constant $k_D$ can be related to the molecular polarizability, with polarizability decreasing with increasing stiffness. The Drude particle gets assigned a small mass, so that it can be integrated with other particles using standard equations of motion. This simplification circumvents the issue of treating polarizability effects on the quantum-mechanical level, while still being able to reproduce spatial variations in polarization. Drude particles do not participate in excluded volume interactions, and thus the only forces acting on them are electrostatic in nature. We note that our implementation is different from those typically used in atomistic or more fine-scale coarse-grained models, where the Drude particle is typically thermostatted independently at a low temperature, enabling the polarizability to be estimated with the classical expression $\alpha = \delta q_D^2/k_D$. Since our charges are distributed over a unit Gaussian, this expression does not apply, and we have parameterized our effective dielectric constant as a function of the various parameters of the Drude oscillators ($q_D, k_D$, and the spread of the Gaussian, $\sigma$), which is presented below in Example Systems.

\subsection{Dynamic bonding and Lewis acid-base pairs}
\label{subsec:Dynamic bonding}

In addition to static bonds, which are initialized at the beginning of the simulation, and remain unchanged throughout its course, MATILDA.FT allows for dynamic bonds to be formed between particles during the simulation run. Dynamic bonds are created or destroyed based on the user-defined acceptance criterion. Currently, Metropolis-Hastings acceptance criterion is used, with an optional shift in the reaction energy. The energy of the bond is calculated based on the extension of the harmonic spring assigned to the pair of bonded atoms.

Dynamic bonds are created between a donor and an acceptor particle. Whether a particle acts as a donor or an acceptor is specified in an external text file, which maps these roles onto particle indices. To make the process computationally efficient while executing on a GPU, MATILDA.FT utilizes a dedicated neighbour list style. Only the donor particles are allowed to initialize the process of bond making and breaking. They store information only about the acceptor particles present in their vicinity. Two separate lists keep track of bonded and free donor particles. During each bonding/unbonding step, two kernels are launched in a randomly chosen order. One kernel attempts to break some of the existing bonds, while the other creates new ones. These kernels are only dispatched using the indices of relevant donor particles - free donors for the bonding kernel, and bonded donors for the bond-breaking one.
Dynamic bonds can also be used for simulating induced dipoles, for example the Lewis acid-base pairs. This requires only a minimal change in the input script, namely, specifying the magnitude of the charge to be assigned to each of the bonding partners. Partial charge of opposite sign gets assigned to donors and acceptors upon binding, so the net charge of the system remains constant.

\subsection{Hydrodynamics using DPD Thermostat}
\label{subsec:dpd-method}

Random noise is introduced into the simulation to account for the lost degrees of freedom that are integrated out during the coarse-graining process. Although Brownian Dynamics can be used to capture the behaviour of polymer solutions, the resulting equations of motion do not satisfy the fluctuation-dissipation theorem. Thus, they do not reproduce correct hydrodynamics. An alternative method, available in MATILDA.FT, which obeys Navier–Stokes equations, is the Dissipative Particle Dynamics (DPD) thermostating. A brief overview of the method is provided below. For more detailed description, we refer the reader to \cite{groot_role_1999}.
In this method, three types of forces act on particles. A conservative force $\mb f^c$, frictional (dissipative) force $\mb f^d$, and a random force $\mb f^r$. The conservative force corresponds to the gaussian repulsions, and to the electrostatic interactions. The dissipative and random forces are pairwise additive and thus the local momentum is conserved. The dissipative force on particles i is given by Eq. \ref{eq:f_d},
\begin{equation}
\label{eq:f_d}
\mb f_{i}^{d} = -\frac{1}{2}\frac{\sigma^2}{k_bT}\sum_j(\omega(|\mb r_{ij}|))^2(\mb{v}_{ij} \cdot \mb{r}_{ij}) \, \mb{r}_{ij},
\end{equation}
whereas the random force is given by
\begin{equation}
\label{eq:f_r}
\mb f^r_i = \sigma\sum_j(\omega(|\mb r_{ij}|))\gamma_{ij} \frac{1}{\sqrt{\Delta t}} \hat{\mb r}_{ij},
\end{equation}
where $\hat{\mb r}_{ij}$ indicates a unit vector in the direction of $\mb r_{ij}$.

As DPD involves the use of range-limited forces, it requires a neighbour list. Since the interactions are pairwise-additive and symmetric, only a "half" list is needed, where each particle only stores the information about the particles with the index lower than its own. This reduces the amount of required computation, and ensures optimal GPU thread utilization. The DPD thermostat should be used with the Velocity Verlet integrator. Code performance is affected by the choice of the value of $\sigma$, and the time step. Strategies for choosing the values and the reasoning behind them is outlined in the documentation.

\section{Code structure, extensibility, and flexibility}
\label{sec:code structure}

This section will give a brief overview of the organization of the code and how its different parts cooperate with each other. 

\subsection{Input Script}

The first step of the simulation is selecting the method to be used, either TILD or FT. This is passed as a command line argument when the program is called. Using \texttt{./MATILDA.FT -particles} will run the TILD simulation, whereas \texttt{-ft} option will initialize an FT run. Additional command line arguments, like the name of the input script to be used, are described in the documentation.
Two files are required for a TILD simulation. The main input file is responsible for setting up simulation parameters, such as the dimensionality of the system, size of the simulation box, grid density, time step size, and the number of time steps to perform. 
The input script also defines the interaction potentials between selected atom types (Potentials), and the parameters used for calculating electrostatic forces - Bjerrum length and the charge spreading length (currently uniform for the entire system, will be allowed to vary between different types of particles in the future release). The same file also specifies particle groups, along with the corresponding neighbour lists, integrators, and additional forces to operate on selected particles.

The second (data) file provides the initial positions of the particles, their types, and the molecules they belong to. This data file also initialize the static bonds and angles used in the simulation.
Currently, the initial atom configuration can be read either from the LAMMPS data file (in angle or charge style) or from a GSD file.

\subsection{Code Organization}

\begin{figure}
    \centering
    \includegraphics[width=0.5\textwidth]{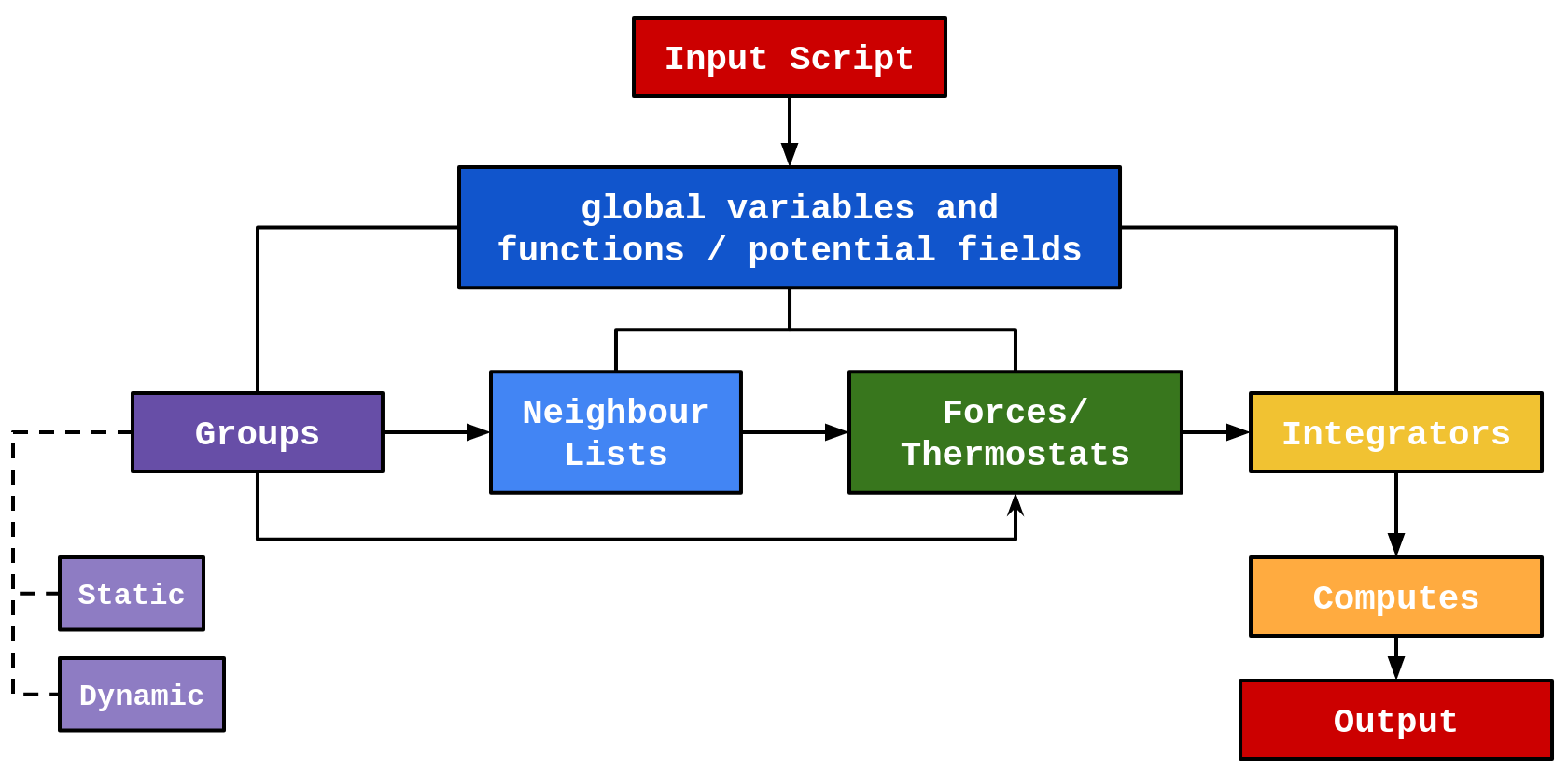}
    \caption{Schematic outline of code structure}
    \label{fig:code_outline}
\end{figure}

The code takes advantage of C++ object oriented programming approach. It is divided into classes, which interact with each other and exchange data as needed. Each class is responsible for handling a particular functionality. The base class serves as an interface used to interact with other parts of the code. Then specialized sub-classes are derived from the base class, to provide a specific functionality. This organization makes extending the code to include new functionalities a relatively easy and straightforward process, with simple integration of the new components into the existing code \ref{subsec:extensibility}. For example, the NeighbourList class is responsible for constructing and storing the neighbour list for the selected group of particles. This neighbour list is then used by the additional forces (created as a subclass of the ExtraForce class) to accelerate the operations performed on this group. Depending on the nature of the additional force, specialized neighbour lists can be used, in order to further accelerate the performance.
A diagram of class code organization in shown in Fig. \ref{fig:code_outline}. Below, we provide a brief description of selected classes. The outline of the code structure, along with the detailed description of all class functionalities and options are provided in the documentation.

\subsubsection{Global variables space}
The global variable space holds the main data structures used in the simulation. It stores the global arrays containing particle types, positions, forces acting upon them, velocities, and static bonds, arranged according to the particle ID. Reading of the input script is also handled at this level. Before the beginning of the simulation, these arrays are initialized and then periodically updated by other classes during the time-stepping process.
In the future release these structures will be placed in a separate Box class, to closely resemble to organization of the FT branch of the code.

\subsubsection{Group Class}

The \textbf{Group} base class provides data structures which store indices of the member particles. It sets device-specific variables (BLOCK and GRID sizes) that are used in kernel calls dispatch on the group particles. All forces and neighbour lists in MATILDA.FT operate on specific groups. Pointers to each group object are stored in a globally accessible vector array. Each group is assigned a unique name, which is used to pass its pointer to their classes. Groups can be static or dynamic. Static groups are initialized at the beginning of the simulation and their content remain unchanged over the simulation course. Dynamic groups, on the other hand, periodically check and update their members, based on the specified membership criterion.
A special group, named \textit{"all"}, is initialized by default at the beginning of the simulation, and contains all particles in the simulation box.
Currently, two static group types are available - grouping by particle \textbf{type} or by its global \textbf{id}. Type-based groups collect all the particles with the same type (as specified in the input.data). Id-based groups require the user to provide an external plain text file which has contains the indices of the particles to be included in the group.
Currently available dynamic group style, ``regions'', allows the user to define a separate region in space (along all or only specific axis). Particles found within that region get assigned to the group.
The can be easily extended to incorporate user-defined groups, see Section \ref{subsec:extensibility} below.

\subsubsection{Neighbour List Class}

Like all operations in MATILDA.FT, neighbour lists act on groups. For the purpose of neighbour list building, the simulation box is assigned a grid which divides it into discrete, non-overlapping cells. Interfacing with a specific group of particles and the division of the simulation box into non-overlapping cells is handled by the \textbf{NList} base class.

Building up on this basic functionality, more specialized sub-classes of neighbour lists are created. The simplest one, \textit{distance}, is the "full" neighbour list, which stores the entire information about particle proximity, and thus double-counts each pair of neighbours. A slightly more elaborate sub-class, \textit{half\_distance}, stores only half of this information. More specifically, each particle only keeps track of the particles which have a lower id than their own. This method also avoids unnecessary calculations when pairwise interactions are present. Otherwise, a branching statement needs to be included, to perform the computation on only one member of the particle pair. Alternatively, the same calculation is performed for both members. Unfortunately, thread divergence is not permitted in the GPU model of execution, and thus a branching statement forces all the threads to wait until the other ones have competed the first branch. This waste of resources is avoided by using an appropriate neighbour list.

A dedicated neighbour list has been designed specifically to support the dynamic bonding functionality. Briefly, only the donor particles, which initialize the bonding, are assigned the neighbours. These neighbours are then filtered to only include the acceptor particles, as only the donor-acceptor combination can create a valid bonded pair. By pre-calculating the possible pairs at this step, these checks can later be avoided in the kernel calls (preventing possible thread divergence).

\subsubsection{ExtraForce Class}

In addition to electrostatic and repulsive interactions, selected groups of particles can be subject to additional user-defined forces. These are specified in the input script using the \textit{extraforce} command. The \textbf{ExtraForce} base-class is responsible for assigning the force to the specific group of particles, and ensuring it is applied to this group at the specified time-steps (wither each step or user-defined frequency). In addition, some range-limited forces require a neighbour list to restrict the search space only to the particles present within the specific range. Currently available forces are:
\begin{itemize}
    \item \textbf{Wall} - which enables the particles to be confined within a specific region or to simulate surface interactions. The user can chose from available wall-particle potentials, or specify their own form of interaction, by extending the source code.
    \item \textbf{Langevin} - Adds random noise to the selected group of particles. Can be used with Velocity Verlet integrator to simulate Brownian dynamics.
    \item \textbf{Midpush} - Adds a force to push the selected group of particles towards the center of the box along a specified axis.
    \item \textbf{DPD} - Dissipative Particle Dynamics. This subclass of ExtraForce provides an alternative way to introduce random noise into the simulation, and should be used along with the Velocity Verlet integrator. In contrast the Langevin thermostat, however, it is pair-wise additive and conserves local momentum. Thus it capable of correctly reproducing hydrodynamic behaviour of the system. Since the force acts over a limited range, a neighbour list needs to be constructed for the particles of interest.
    \item \textbf{Lewis} - This additional forces can be used to introduce dynamic bonds in the simulation. Whereas static bonds are initialized at the beginning of the simulation, and remain unchanged, dynamic bonds can be formed and broken according to the specified acceptance criterion. This force requires a specialized neighbour list (\textit{bonding}) which has been designed to optimize the required computations.
\end{itemize}

More details about the ExtraForce class can be found in the documentation.

\subsubsection{Compute Class}
The \textit{Compute} class is responsible for performing on-the fly calculations of properties of the system. This enables the user to monitor the evolution of the system in real time and also saves time spent on post-processing.

\begin{itemize}
    \item \textbf{Average Structure Factor} $\langle S(k)\rangle$ - this compute provides the information about the average static structure factor of the particle system. The static structure factor, given by
    \begin{equation}
    \label{eq:sk}
        S(k) = \frac{1}{N} \langle \hat{\rho}_k \hat{\rho}_{-k} \rangle,
    \end{equation}
    is defined as the correlation function of the system density represented in the Fourier space. The density is given by $\rho(r) = \sum_{i = 1}^{N} \delta(r-r_i)$, and in Fourier space it becomes  $\hat{\rho}_k = \sum_{i = 0}^{N} e^{i k \cdot r_i}$ \cite{zhang_concept_2016}.
    It performs the calculation and writes the data to an external file according to user-specified frequency.
    
    \item \textbf{Chemical Potential} $\mu$ - this compute can be used to calculate chemical potential of the given species in the system using a chain deletion method. User provides a range (based on global molecule ID) of molecules to operate on. These particles are removed at random from the system, and the change in energy upon removal is calculated. This data is written to an external file.

\end{itemize}

\subsubsection{Integrator Class}

In order to solve the equations of motion and propagate the particle coordinates in time, numerical integration is required. In MATILDA.FT, three different numerical algorithms are available, and are briefly described below:

\begin{itemize}
    \item \textbf{Velocity-Verlet (VV)}. Needs to be coupled with additional thermostat. Available thermostats include Langevin noise or Dissipative Particle Dynamics, which of which are part of the ExtraForce class.
    \item \textbf{Euler-Maruyama (EM)}. Generates the thermal noise internally during the update and serves as the simplest stochastic integration scheme to implement.
    \item \textbf{Grønbech-Jensen and Farago (GJF)}. Generates thermal noise internally during the update, and We find that this algorithm allows time steps up to 10x larger than the EM algorithm with no loss of accuracy or stability.
\end{itemize}

\subsection{FTS Implementation}

\subsubsection{Class structure}
\begin{figure}
    \includegraphics*[width=\columnwidth]{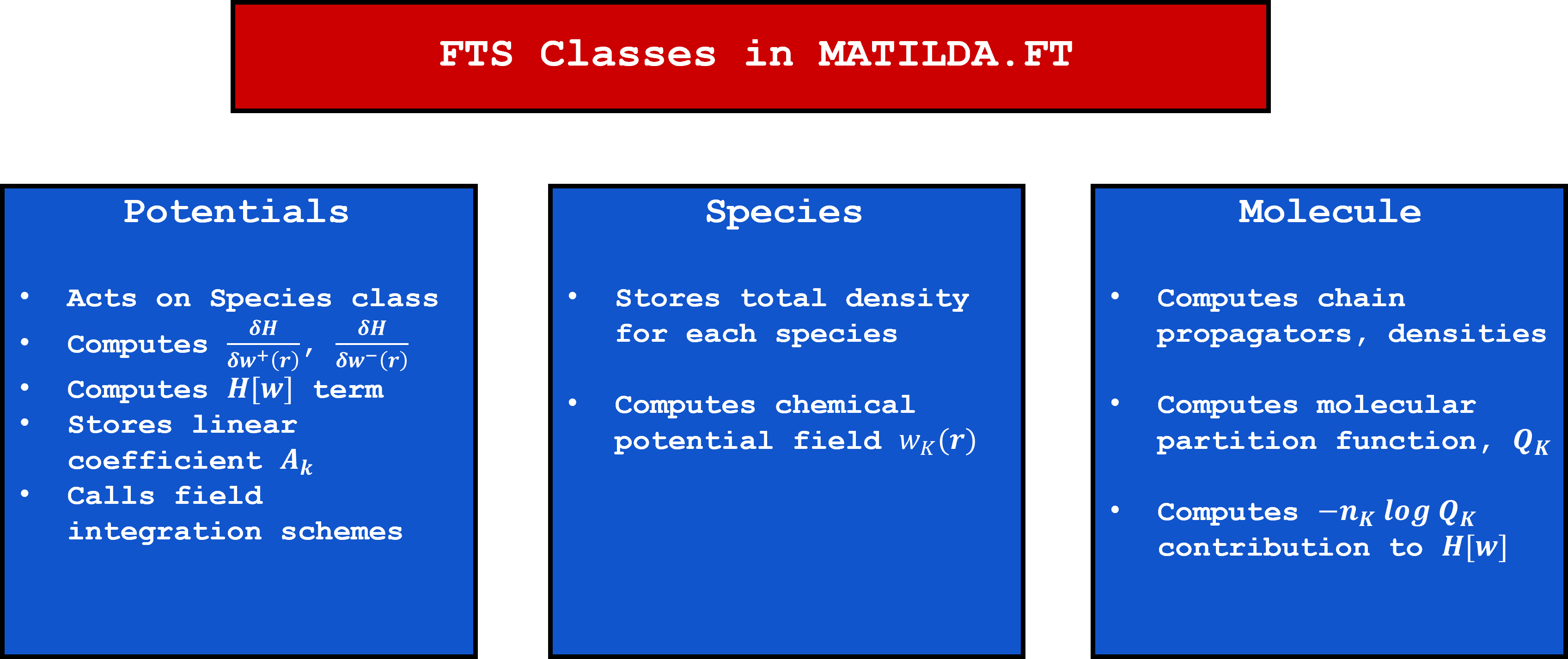}
    \caption{Basic actions and roles of the FTS Classes in MATILDA.FT.}
    \label{fig:FTclasses}
\end{figure}

As the FTS branch was begun more recently, many planned features are still in development. There is also a difference in class organization between the older (TILD), and the newer (FT) branches. The classes that comprise an FT simulation are more tightly integrated and the scope of object-oriented organization is larger, as compared to the TILD branch which still uses global variables.

\begin{figure}
    \begin{center}
    \includegraphics*[width=0.8\columnwidth]{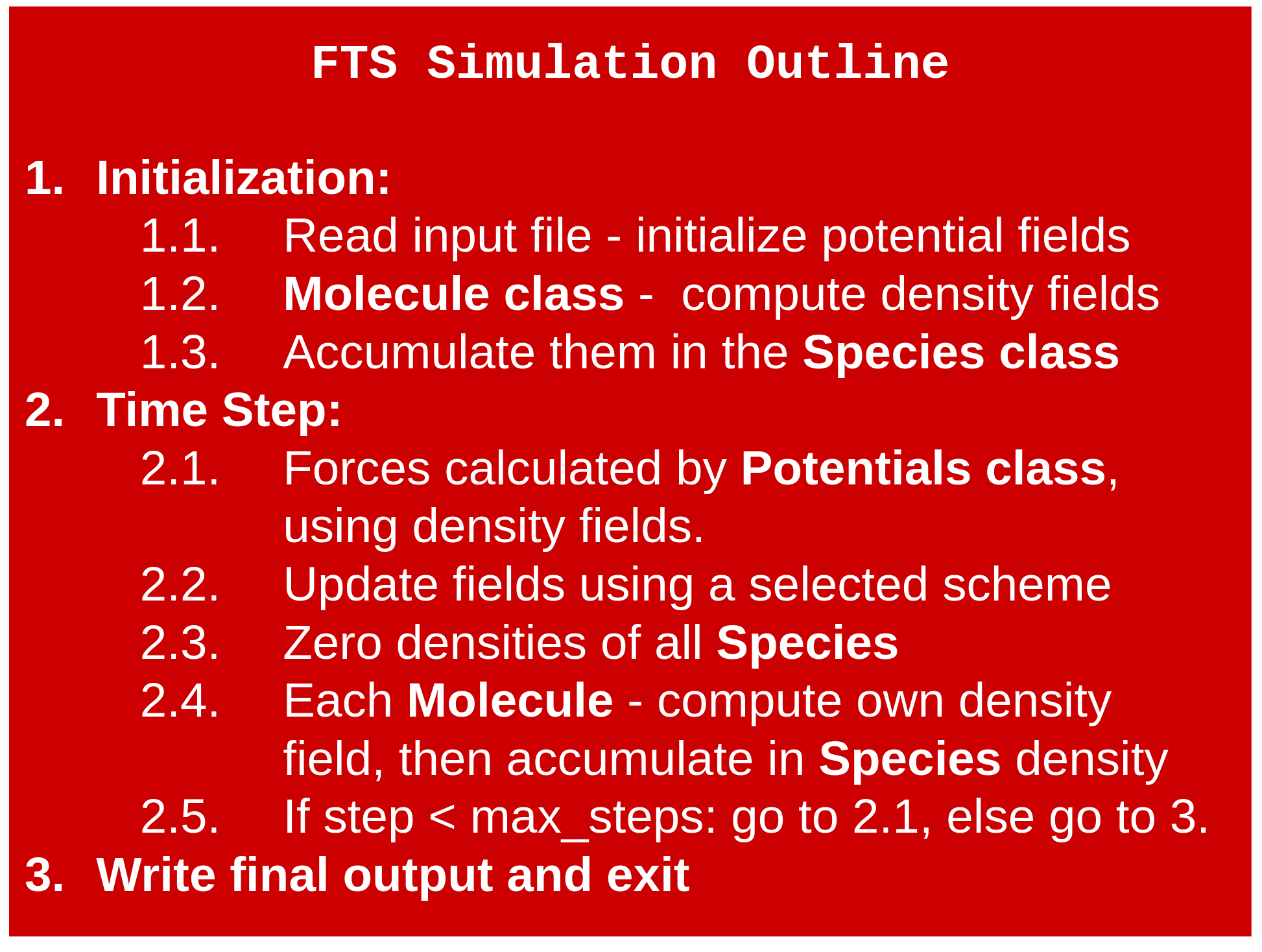}
    \end{center}
    \caption{Outline of an FTS simulation as implemented in MATILDA.FT. The termination condition could be convergence to within a prescribed tolerance in SCFT or reaching the maximum desired number of time steps in a CL simulation.}
    \label{fig:FToutline}
\end{figure}

A field-theoretic simulation lives in an \textbf{FTS$\_$Box} class, which contains three key classes: \textbf{Potentials}, \textbf{Molecules}, and \textbf{Species} (see Figure \ref{fig:FTclasses} for graphical outline of FT branch organization). The \textbf{Potentials} class performs all of the functions that are related to the the various non-bonded interactions, including updating the potential fields associated with a particular interaction. The densities that show up in the effective forces are taken from the \textbf{Species} class, which serves as a container for these densities. {\bf Species} generates the unsmeared chemical potential fields, $\mu_{K,c}(\mb r)$, by looping over the interaction potentials and accumulating the relevant potential fields. Next, the \textbf{Molecules} class takes these potential fields, applies any density smearing that may be necessary, and computes both the center and smeared density fields. The {\it smeared} density fields are then accumulated into the relevant \textbf{Species}  class. The general flow of the code is summarized in Figure \ref{fig:FToutline}.

\subsubsection{Field update schemes}
Currently, two schemes have been implemented to update potential fields, in order to evolve in time equations of motion such as Eq. ~\ref{eq:CL}.
The straightforward explicit Euler-Maruyama (EM) integration scheme discretizes the equation in time and uses the forces at the current time to estimate the field configurations at a future time as
\begin{equation}
    w^{t+\delta t}(\mb r) = w^{t} + \delta t\, \lambda_w F_w^t(\mb r) + \sqrt{2\delta t \lambda_w} \zeta_t(\mb r),
    \label{eq:EM}
\end{equation}
where $\delta t$ is the size of the time step and $\zeta_t(\mb r)$ is purely real Gaussian noise with unit variance that is uncorrelated in both space and time. As mentioned above, simply neglecting the noise term converts this algorithm to one that drives the system to a mean-field solution. 

The other algorithm that is implemented is a 1st-order, semi-implicit updating scheme, that has been shown to allow for time steps significantly larger than allowed by the EM scheme\cite{lennon2008numerical, ceniceros2004numerical, villet_efficient_2014}. In this approach, one derives an approximate expression for the force $F_w^{t, lin}(\mb r)$ that is linear in the potential field. In real-space, these expressions take the form of a convolution
\begin{equation}
    F_w^{t,lin}(\mb r) = \int d\mb r' \; A_w(\mb r - \mb r') \, w(\mb r'),
    \label{eq:Flin_r}
\end{equation}
where $A_w(\mb r)$ is the linear coefficient. As a result of this convolution, the 1S updating scheme is most effectively handled in Fourier space where we have
\begin{equation}
     F_w^{t,lin}(\mb k) = A_w(\mb k) \, w(\mb k).
    \label{eq:Flin_k}
\end{equation}
To affect the semi-implicit scheme, Eq.~\ref{eq:EM} is written in Fourier space and modified by subtracting the linear term at $t+\delta t$ and adding it at $t$ giving,
\begin{equation}
\begin{array}{rcl}
    w^{t + \delta t}(\mb k) & = & \delta t\, \lambda_w \left[ F_w^t(\mb k)  + A_w(\mb k) \, w^t(\mb k) - A_w(\mb k) w^{t+\delta t}(\mb k) \right]
    \\ && \\
    && + w^t(\mb k) + \sqrt{2\delta t \lambda_w} \zeta_t(\mb k).
\end{array}
\label{eq:pre1S}
\end{equation}
We note that $\zeta_t(\mb k)$ is generated as a spatially uncorrelated noise field in real-space that is explicitly Fourier transformed. Equation \ref{eq:pre1S} can be readily solved for the field at $t+\delta t$ giving
\begin{equation}
    w^{t + \delta t} = \frac{w^t + \delta t\, \lambda_w \left[  F_w^t  + A_w \, w^t \right]+ \sqrt{2\delta t \lambda_w} \zeta_t}
    {1 + \delta t \lambda_w \, A_w},
    \label{eq:1S}
\end{equation}
where we have suppressed the wavevector dependence for brevity.

The functional form of the linear coefficients $A_w$ generally contain one or two contributions that have a stabilizing effect on the time integration\cite{lennon2008numerical, ceniceros2004numerical}. The first arises from the terms that are quadratic in the fields in $\mc H$ (e.g., the first two lines in Eq.~\ref{eq:H}); this term is included for every type of interaction potential. During the initialization of an FT simulation, the \textbf{Potentials} class adds this relevant term to $A_w$. The second contributions are the linear approximates of the density operators, which involve convolutions of Debye functions with the potential fields; these contributions are handled by the {\bf Molecules} class during initialization.

\subsubsection{Molecule Types}
Currently, the only implemented molecule type is a linear, discrete Gaussian chain with an arbitrary number of blocks. This class handles the calculation of the chain propagators, and during initialization the code automatically checks whether the molecule is symmetric to avoid calculating the complimentary propagator if possible. The other key step taken in the initialization is to accumulate the relevant Debye-function-like contributions to the linear coefficients associated with each potential. From previous studies \cite{lennon2008numerical, ceniceros2004numerical}, not all terms that show up in the precise linear expansion of the force are stabilizing; to that end, we do not include the Debye terms in the $w_{AB}^{(-)}$ field, but they are included in the $w_+$ and $w_{AB}^{(+)}$ fields.

\subsection{Extensibility}
\label{subsec:extensibility}

MATILDA.FT has been designed to be easily extensible by other users, according to their specific needs. The process of expanding the source code is simplified thanks to the division into classes, consistent with C++ philosophy of Object Oriented Programming. Due to this organization, the user needs only to understand how the base class interfaces with other parts of the program, and does not need to rewrite the entire logic. Simple functionalities can easily be added by inheriting the capabilities of provided base classes or modifying the existing once.

\begin{lstlisting}[language=C++,caption={Code extensibility example},captionpos=b, label=extension, style = code]
/* kernel function to update group members
    based on their position and type */
__global__ void d_CheckGroupMembers(
    const float* x, //position array
    thrust::device_ptr<float> d_wall_data,
    const int n_walls, // number of walls
    thrust::device_ptr<int> d_all_id,
    const int ns, // group size
    const int Dim, // Dimensionality
    const int* tp){
    // tp[] array stores particle types
    int list_ind = blockIdx.x * blockDim.x + threadIdx.x;
    if (list_ind >= ns)
        return;
    int ind = list_ind;
    for(int i = 0; i < n_walls; ++i){
        int j = int(d_wall_data[3 * i]);
        float low = d_wall_data[3 * i + 1];
        float high = d_wall_data[3 * i + 2];
        float xp = x[ind * Dim + j];
        d_all_id[ind] = ind;
        if (xp>=low && xp<=high && tp[ind]==1)
            // additional type check
            d_all_id[ind] = ind;
        else
            d_all_id[ind] = -1;
    } // i < n_walls
\end{lstlisting}

This simplicity is demonstrated in Listing \ref{extension}, where a new group is produced. It is based on the dynamic group which only includes particles present in the specific region. Here, in addition to spatial criterion, only particles of a particular type (here 1) are considered group members. Here, only the modified GPU kernel is shown for brevity. Two changes to the original code were introduced to achieve new functionality. The first one on line 10 (an additional parameter in the function call), and another on line 22 (type check). This file can later be incorporated into a new subclass of the \textbf{Group}, compiled with the rest of the source code, and will work seamlessly in the simulation. Included with the source-code is also a pre-made make file which makes compilation of additional components easy, only requiring that the be added to the source list. The full example, along with a step-by-step explanation, and make file description is available in the \textit{examples/extend} folder in the GitHub repository.

\section{Example systems}
\label{sec:examples}

\subsection{Coacervate}

\begin{figure}
    \centering
    \includegraphics[height=0.3\textheight]{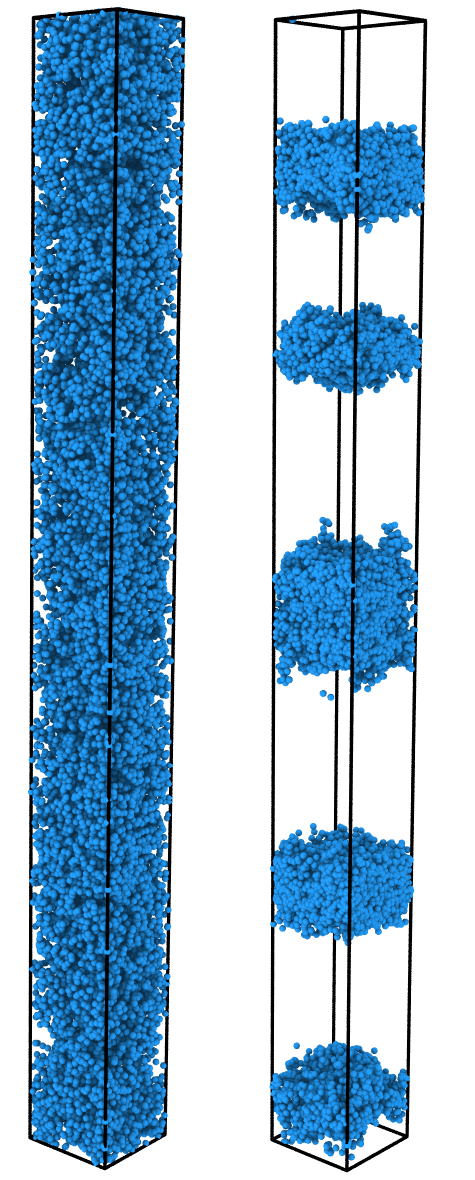}
    \caption{Initial (left) and final (right) snapshot from the simulation of coacervating binary mixture. This simulation is a particle-based implementation of the model considered previously as a field-theory by one of us\cite{riggleman2012investigation} with a dimensionless excluded volume parameter $B = 0.05$ and dimensionless Bjerrum length $E = 10000$. }
    \label{fig:coacervate}
\end{figure}

In this example, a small system consisting of the total of 434 molecules, each with a degree of polymerization, $N = 82$, has been simulated. Half of these molecules carry positively charged monomers, with the other half having each monomer with negative charges of the same magnitude. No explicit solvent is present. The system starts in a random, homogeneous phase and over the course of simulation phase separates into polymer-rich and polymer-depleted regions as coacervation occurs. The snapshots from the beginning (left), and the end (right) are shown in Figure \ref{fig:coacervate}.  On the Nvidia Quadro RTX 5000 GPU, the simulation took 1399 seconds to perform 2,000,000 time steps, with the resulting speed of 1429.6 ts/sec.
This example can be found in the GitHub repository, in the \textit{examples/Coacervate} directory. The full movie of the time evolution of the system is available in the supporting information.

\subsection{Hydrodynamics using DPD integrators}

\begin{figure}
    \centering
    \includegraphics[width=\columnwidth]{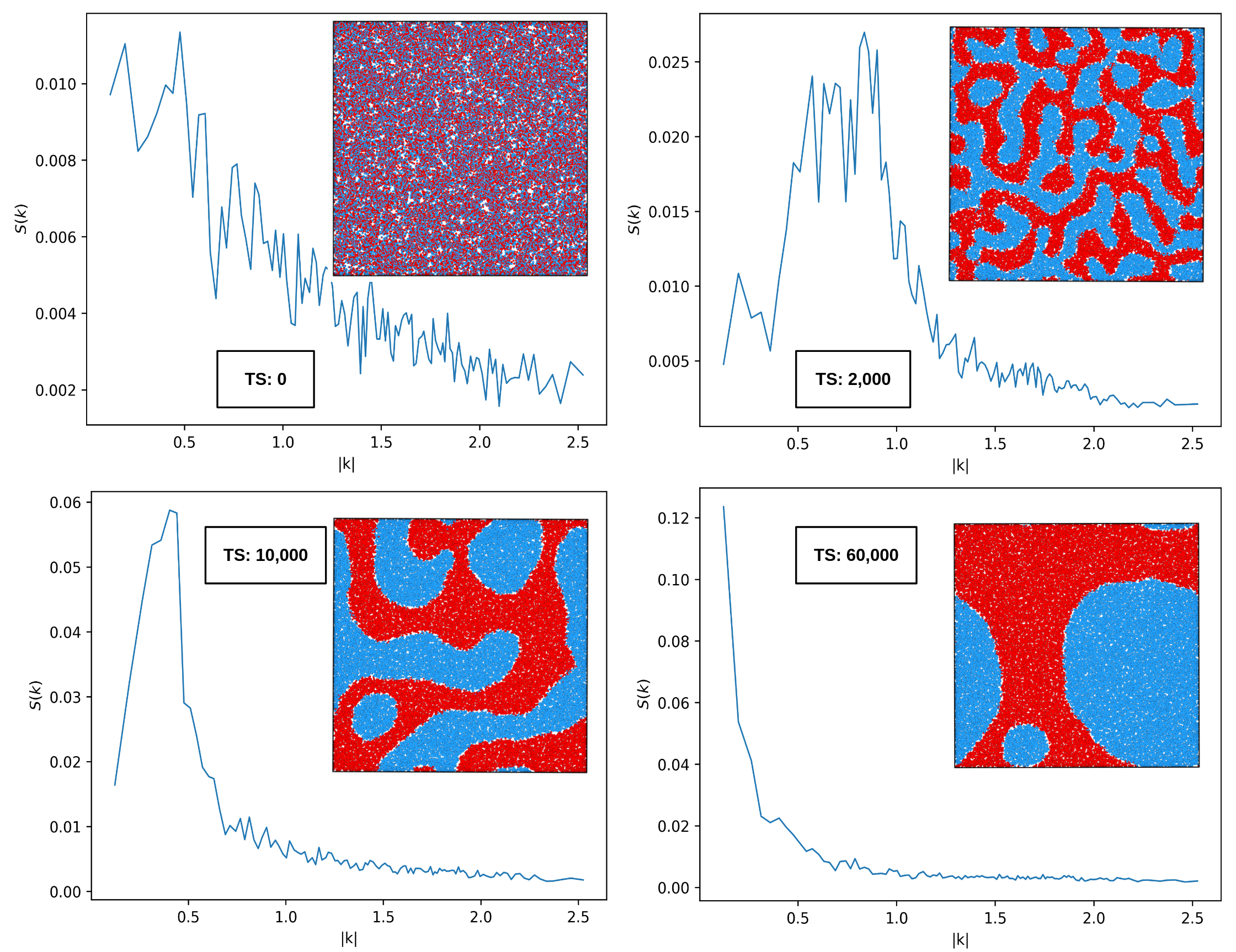}
    \caption{ Static structure factor $S(k)$ calculated on the red species undergoing spinodal decomposition at various time steps (TS) using the DPD integrator. The insets show representative snapshots of the system at the same times.}
    \label{fig:dpd}
\end{figure}

To demonstrate the effect of using DPD thermostat, we simulated a simple binary mixture of uncharged particles in a two-dimensional box. The value of $\kappa$ in the Helfand potential has been chosen to be equal 1 and $\chi_{AB} = 5$ to induce strong phase separation. The value of $\sigma$ was arbitrarily set to 0.5, with $r_{cutoff} = 1.0$. The neighbour list update frequency was set to every 6 time-steps, and $r_{skin} = 2.5$. On the Nvidia Quadro RTX 5000 GPU, the simulation took 5259 seconds to perform 1,200,000 time steps, with the resulting speed of 228.18 ts/sec. In Figure \ref{fig:dpd}, the insets show representative snapshots from the course of the simulation along with the corresponding structure factor for the red species in the main figure.

\subsection{Polarizable diblock co-polymer with explicit solvent}

In this example, 1330 diblock co-polymer chains with $N = 74$, and blocks of equal size, were simulated in explicit solvent. Both the solvent and polymer chains are were made polarizable through the use of classical Drude oscillators. Since in this system only the polymer concentration of the condensate was of interest, all simulations were started with the polymers in a dense ``slab'' configuration. In this configuration all particles are biased to migrate towards the middle of the simulation box, creating a homogeneous, dense polymer phase. During the production run this bias is removed, and the slab is allowed to expand.

\begin{figure}[htbp]
    \centering
    \includegraphics[width=\columnwidth]{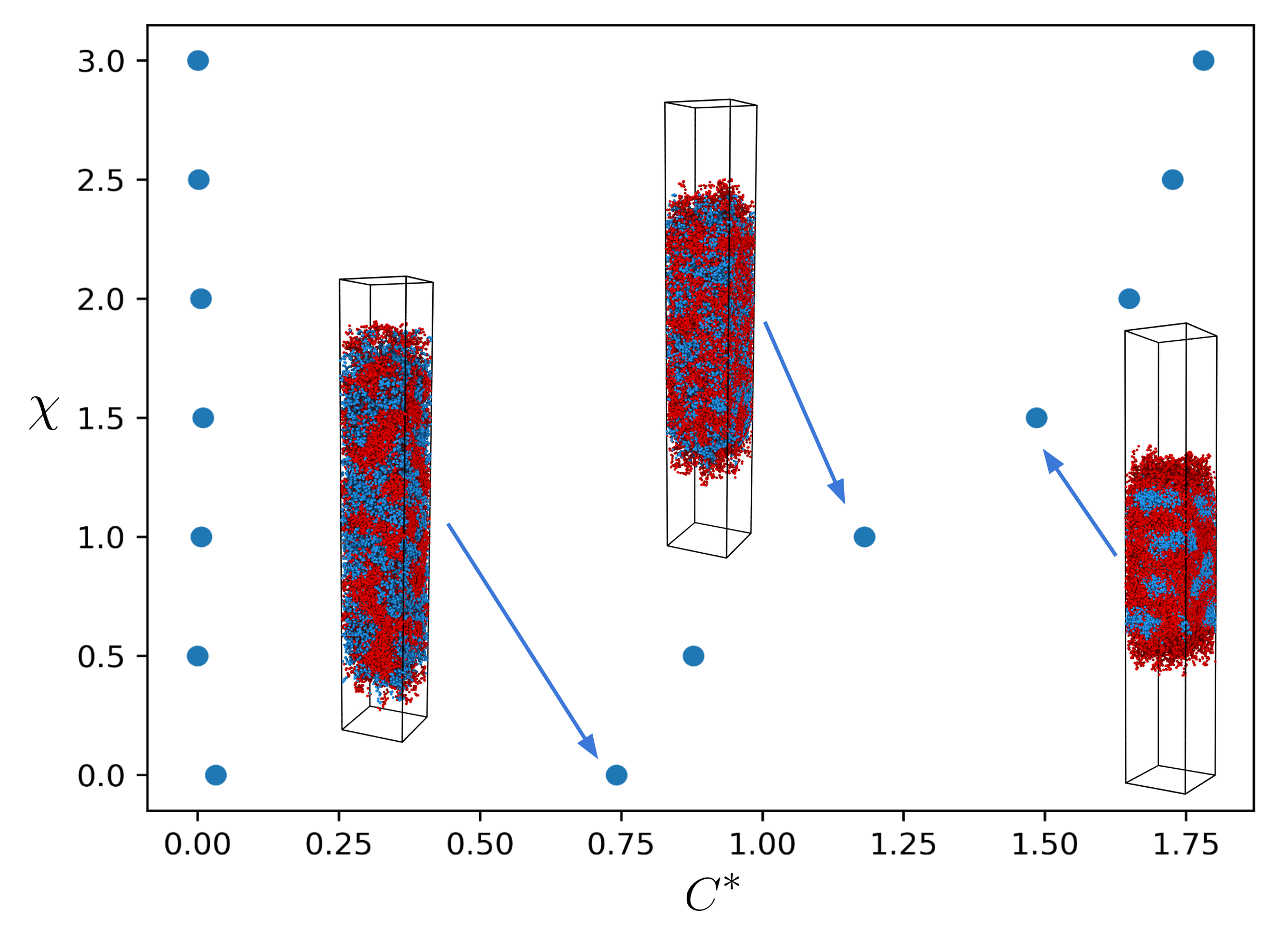}
    \caption{Plot of the density obtained in the dilute and dense phase as the value of $\chi$ between the monomers and solvent is varied. Included are also renders of the three-dimensional structure of the system corresponding to the selected data points. Positively charged monomers are displayed in red, while negatively charged ones are colored blue.}
    \label{fig:ac}
\end{figure}

\begin{figure}[htbp]
    \centering
    \includegraphics[width=0.95\columnwidth]{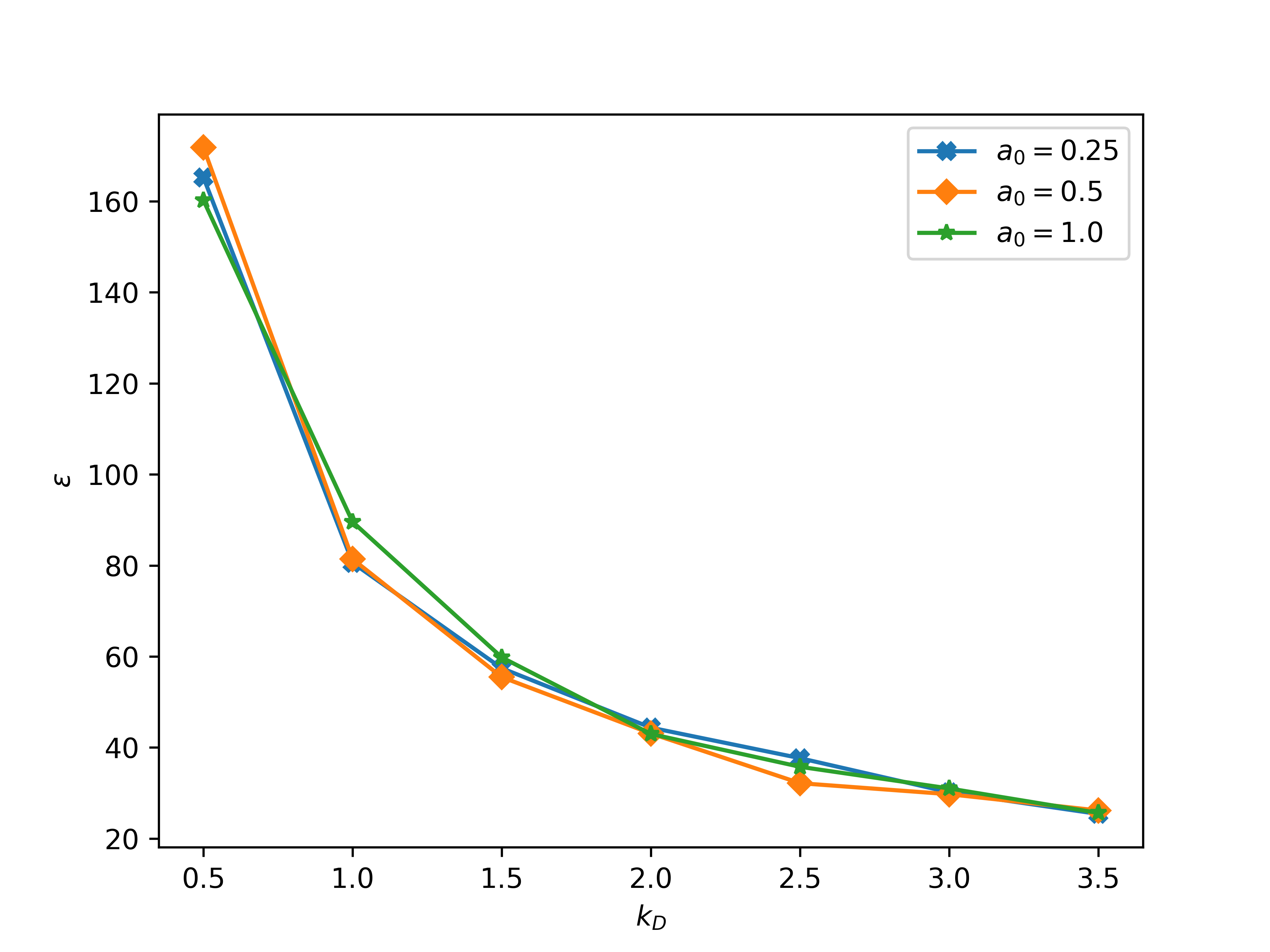}
    \caption{Calculated values of dielectric constant for the solvent molecule, as a function of charge spreading length, $a_0$ and the spring constant of the Drude oscillator, $k_D$}
    \label{fig:water}
\end{figure}

The parameters for the Drude oscillator attached to the solvent molecules have been chosen as to reproduce the dielectric constant of water. In this way, the Bjerrum length can be set to the value it has in the vacuum, and dielectric screening is then emergent from the polarizable solvent. The calculated dielectric constant for chosen combination of parameters are shown in Figure \ref{fig:water}. The dielectric constant of water is around 78.4, so that optimal choice of parameter corresponds to $a_0 \approx 0.5$ and $k_D \approx 1.0$ In Figure \ref{fig:ac} we show the plot of the reduced concentration $c^{*}$ of the dense and dilute phases phase, and the corresponding value of $\chi$ between the polymer monomers and the solvent molecules. We also include corresponding snapshots of final structure of the system.

\subsection{Liquid crystals}
\begin{figure}[ht]
    \includegraphics*[width=\columnwidth]{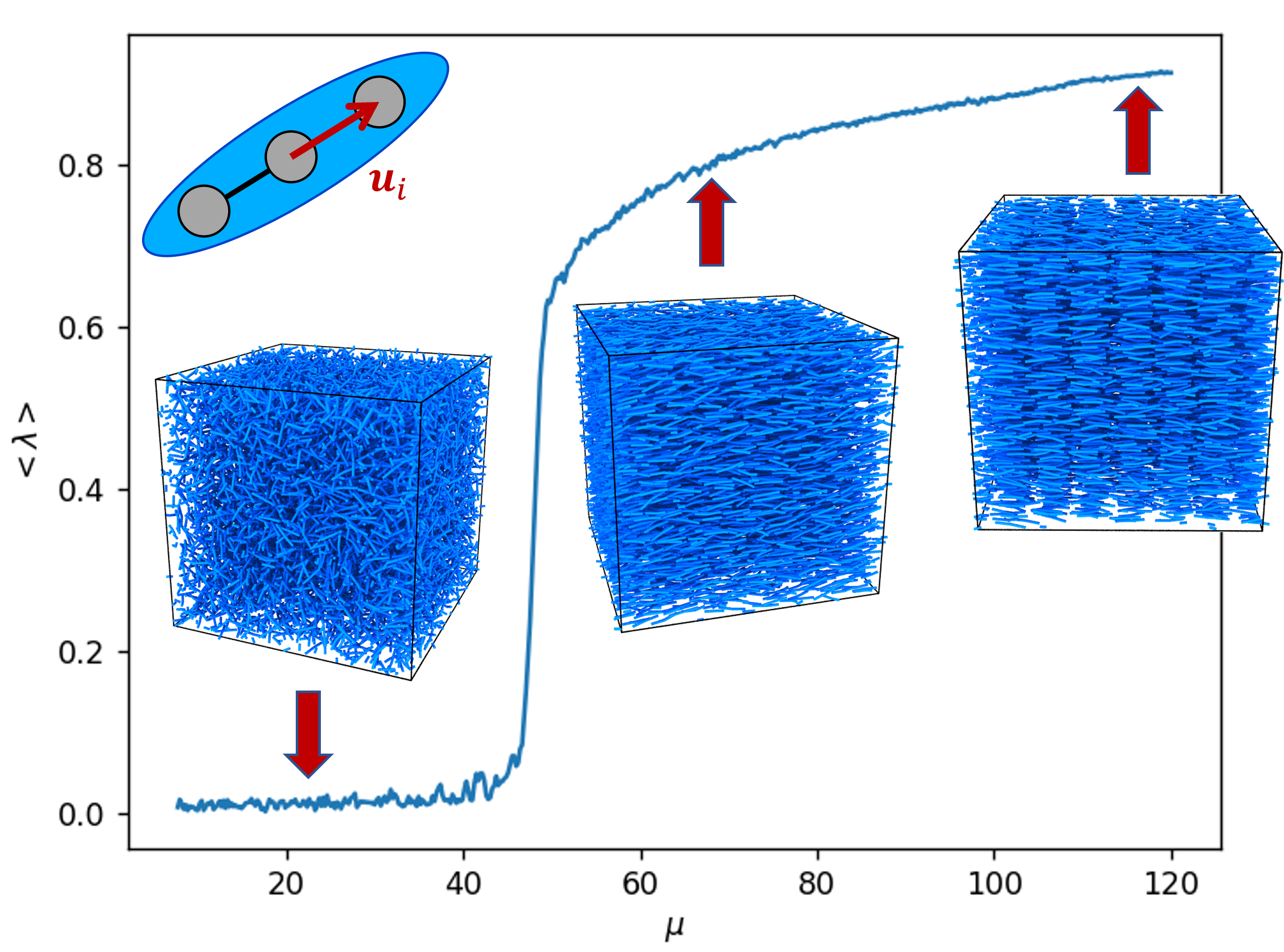}
    \caption{Liquid crystalline order parameter as a function of the strength of the Maier-Saupe parameter $\mu$ calculated during a simulation where $\mu$ was continuously ramped throughout.}
    \label{fig:LC}
\end{figure}

A simple model of a pure liquid crystal was simulated where the mesogen was discretized into three interaction sites with the Maier-Saupe (MS) interaction taken from the center of the mesogen, see the inset in Figure \ref{fig:LC}. Bonds between adjacent liquid crystal sites used a force constant $k_b = 100$ and equilibrium distance $1$, and the orientation was maintained with a worm-like chain angle potential with prefactor $\lambda = 50$. Finally, an additional Helfand potential was employed to maintain an approximately uniform density with $\kappa = 11$ and the total site density was $\rho_0 = 3$. This combination of the anisotropic molecular shape with a MS interaction taken from its center makes the model similar to the McMillan mean-field model \cite{mcmillan1971simple}. 

Figure \ref{fig:LC} shows the average liquid-crystalline order parameter $\lambda$ that is calculated on the fly as the MS $\mu$ parameter is increased from 0 to 120. $\lambda$ is taken as 3/2 times the largest eigenvalue of the average $\mb S$ tensor. When $\mu \approx 45$, a sharp increase in $\langle \lambda \rangle$ indicates the isotropic-to-nematic transition.

\section{Planned future developments}
\label{sec:future dev}

In this section we present the near-future and long term goals regarding the development of MATILDA.FT. For the TILD branch, the main modification that is currently in progress, will be converting to a fully object-oriented style, to closely follow the design of the newer, FT branch. This will make the two parts of code operate more seamlessly, and will facilitate future modifications and make extensibility easier. Afterwards, we will focus on optimizing existing algorithms, specifically modifying the calculation of repulsive interactions based on Gaussian density.

The long-term goals for the TILD branch include introducing enhanced sampling techniques, such as Umbrella Sampling and Gibbs Ensemble Monte Carlo. This would enable the study of systems that suffer from being stuck for extended periods of time in metastable potential minima, such as polypeptides. We are also planning on expanding the software to be able to perform rigid body dynamics, which can be used to model patchy particles with specific interaction. Finally, we plan to harness the power of using multi-GPU architectures, to aid in simulations with multiple boxes. This feature would greatly facilitate simulations using enhanced sampling techniques like replica exchange, parallel tempering or Gibbs Ensemble MC. A ``quality of life'' feature we plan to implement is internal routines to generate starting configurations so that one could simply specify the polymer architecture, length, and density, and the code provides a starting configuration. Finally, the existing neighbor lists could be used to implement slip-springs to capture entanglement effects\cite{ramirez2015multichain, ramirez2017multi}, which would enable studies of, for example, viscoelastic phase separation.
As the user community grows, we will try to implement features that are in high demand, and would provide useful tools to the scientific community.

We plan to implement several key features for the FT simulation methods that should be available in the near term. While the time evolution equations of the fields described above are presented from the perspective of a complex Langevin (CL) simulation, the CL equations are not yet implemented, and all of the equations of motion are currently noise free. In addition, we plan to implement monomer smearing with unit Gaussians to regularize the models against ultra-violet divergences\cite{koski2013field, riggleman2012investigation, villet_efficient_2014}. Finally, the inclusion of electrostatic interactions, including polar and polarizable polymer monomers\cite{martin_statistical_2016}, is planned in the near future.
Further in the future we plan to implement more exotic polymer arcitectures such as bottlebrush and star polymers. Finally, a variety of nanoparticles, including anisotropic\cite{koski2013field} and polymer-grafted particles\cite{chao2014distribution, koski2015predicting} as both field-based particles and explicit ``hybrid'' particles\cite{sides2006hybrid} are planned for the future.

\section{Concluding Remarks}
\label{sec:summary}

In this paper we presented MATILDA.FT, an open source mesoscale simulation software, which utilizes GPU architecture and primarily a CUDA/C++ programming language to achieve massive parallelism over thousands or millions of particles. The particle-based simulations are primarily designed for highly coarse-grained potentials that are finite on contact and where the particle density will be relatively high so there is a gain in overall efficiency by evaluating the non-bonded interactions using density fields. It has already proven to be able so simulate a vast array of distinct systems and can be easily extended to incorporate new ones, thanks to the object-oriented implementation. As far as we are aware, MATILDA.FT is the first published open-source software to combine both coarse-grained Langevin dynamics and field-theoretic simulation frameworks into a single code base. The code has been written with new users in mind, and its use and basic extensibility do not require specialized programming knowledge. This will make it a valuable to the broad part of the scientific community, providing a powerful computational tool to both theorists and experimentalists across many fields of material science and bio-engineering.
We plan on continuously extending and improving the existing code, and building a user community where scientists can share their ideas and experience. As with many other open-source codes, such as LAMMPS, this approach is extremely beneficial and allows for a faster, and more impactful software development, to keep up with upcoming scientific challenges.

\section{Acknowledgements}

This work used Bridges-2 GPU at Pittsburgh Supercomputing Center through allocation DMR150034 from the Advanced Cyberinfrastructure Coordination Ecosystem: Services \& Support (ACCESS) program, which is supported by National Science Foundation grants \#2138259, \#2138286, \#2138307, \#2137603, and \#2138296. This work was supported by the National Science Foundation through grant MRSEC/DMR-1720530 (R.A.R. and partial Z.M.J.), NSF awards OISE-1545884 (C.T.) and CHE-220375 (C.G.), and the Department of Physics and Astronomy at the University of Pennsylvania (Z.M.J.). The images used in this publication were generated using the free visualization software Ovito version 2.9.0 \cite{stukowski_visualization_2010}.

\section{Data Availability}
The source code is open source and available under the Gnu public license (GPL2) at www.github.com/rar-ensemble/MATILDA.FT. The ``examples'' folder contains all of the input files needed to reproduce the simulations presented herein.

\section{Author Contributions}
{\bf Z.M.J:} writing/original draft (lead); software (supporting); Validation (equal); Visualization (lead); {\bf C.T:} writing/review and editing (supporting); software (supporting); {\bf C.G:} writing/review and editing (supporting); software (supporting); {\bf N.H:} software (supporting); {\bf A.Y:} software (supporting); {\bf R.A.R:} writing/original draft (supporting); review and editing (lead); software (lead); conceptualization (lead); Validation (equal); Visualization (supporting).

\bibliography{matilda.bib,bibio.bib}

\end{document}